\documentclass{pasj01}
\usepackage{natbib}
\usepackage{time}

\RequirePackage{mathpazo}
\RequirePackage[T1]{fontenc}

\Received{yyyy/mm/dd}
\Accepted{yyyy/mm/dd}
\Published{yyyy/mm/dd}

\newcommand{\Msun}{M_{\odot}}
\newcommand{\Mej}{M_{\rm ej}}
\newcommand{\Ye}{Y_{\rm e}}

\begin{document} 

\title{J-GEM observations of an electromagnetic counterpart to the neutron star merger GW170817}

% 1
\author{Yousuke \textsc{Utsumi}\altaffilmark{1}}
% 2
\author{Masaomi \textsc{Tanaka}\altaffilmark{2}}
% 3
\author{Nozomu \textsc{Tominaga}\altaffilmark{3,4}}
% 4
\author{Michitoshi \textsc{Yoshida}\altaffilmark{5}}
% IRSF
\author{Sudhanshu \textsc{Barway}\altaffilmark{6}}
\author{Takahiro \textsc{Nagayama}\altaffilmark{7}}
\author{Tetsuya \textsc{Zenko}\altaffilmark{8}}
% Subaru
\author{Kentaro \textsc{Aoki}\altaffilmark{5}}
\author{Takuya \textsc{Fujiyoshi}\altaffilmark{5}}
\author{Hisanori \textsc{Furusawa}\altaffilmark{2}}
\author{Koji S. \textsc{Kawabata}\altaffilmark{1}}
\author{Shintaro \textsc{Koshida}\altaffilmark{5}}
\author{Chien-Hsiu \textsc{Lee}\altaffilmark{5}}
\author{Tomoki \textsc{Morokuma}\altaffilmark{9}}
\author{Kentaro \textsc{Motohara}\altaffilmark{9}}
\author{Fumiaki \textsc{Nakata}\altaffilmark{5}}
\author{Ryou \textsc{Ohsawa}\altaffilmark{9}}
\author{Kouji \textsc{Ohta}\altaffilmark{8}}
\author{Hirofumi \textsc{Okita}\altaffilmark{5}}
\author{Akito \textsc{Tajitsu}\altaffilmark{5}}
\author{Ichi \textsc{Tanaka}\altaffilmark{5}}
\author{Tsuyoshi \textsc{Terai}\altaffilmark{5}}
\author{Naoki \textsc{Yasuda}\altaffilmark{4}}
% MOA
\author{Fumio \textsc{Abe}\altaffilmark{10}}
\author{Yuichiro \textsc{Asakura}\altaffilmark{10,$\dagger$}}
\author{Ian A. \textsc{Bond}\altaffilmark{11}}
\author{Shota \textsc{Miyazaki}\altaffilmark{12}}
\author{Takahiro \textsc{Sumi}\altaffilmark{12}}
\author{Paul J. \textsc{Tristram}\altaffilmark{13}}
% Observer
\author{Satoshi \textsc{Honda}\altaffilmark{14}}
\author{Ryosuke \textsc{Itoh}\altaffilmark{15}}
\author{Yoichi \textsc{Itoh}\altaffilmark{14}}
\author{Miho \textsc{Kawabata}\altaffilmark{16}}
\author{Kumiko \textsc{Morihana}\altaffilmark{17}}
\author{Hiroki \textsc{Nagashima}\altaffilmark{16}}
\author{Tatsuya \textsc{Nakaoka}\altaffilmark{16}}
\author{Tomohito \textsc{Ohshima}\altaffilmark{14}}
\author{Jun \textsc{Takahashi}\altaffilmark{14}}
\author{Masaki \textsc{Takayama}\altaffilmark{14}}
% core
% normal
\author{Wako \textsc{Aoki}\altaffilmark{2}}
\author{Stefan \textsc{Baar}\altaffilmark{14}}
\author{Mamoru \textsc{Doi}\altaffilmark{9}}
\author{Francois \textsc{Finet}\altaffilmark{5}}
\author{Nobuyuki \textsc{Kanda}\altaffilmark{18}}
\author{Nobuyuki \textsc{Kawai}\altaffilmark{15}}
\author{Ji Hoon \textsc{Kim}\altaffilmark{5}}
\author{Daisuke \textsc{Kuroda}\altaffilmark{19}}
\author{Wei \textsc{Liu}\altaffilmark{1,20}}
\author{Kazuya \textsc{Matsubayashi}\altaffilmark{19}}
\author{Katsuhiro L. \textsc{Murata}\altaffilmark{15}}
\author{Hiroshi \textsc{Nagai}\altaffilmark{2}}
\author{Tomoki \textsc{Saito}\altaffilmark{14}}
\author{Yoshihiko \textsc{Saito}\altaffilmark{15}}
\author{Shigeyuki \textsc{Sako}\altaffilmark{9,21}}
\author{Yuichiro \textsc{Sekiguchi}\altaffilmark{22}}
\author{Yoichi \textsc{Tamura}\altaffilmark{17}}
\author{Masayuki \textsc{Tanaka}\altaffilmark{2}}
\author{Makoto \textsc{Uemura}\altaffilmark{1}}
\author{Masaki S. \textsc{Yamaguchi}\altaffilmark{9}}
\altaffiltext{1}{Hiroshima Astrophysical Science Center, Hiroshima University, 1-3-1 Kagamiyama, Higashi-Hiroshima, Hiroshima, 739-8526, Japan}
\altaffiltext{2}{National Astronomical Observatory of Japan, 2-21-1 Osawa, Mitaka, Tokyo 181-8588, Japan}
\altaffiltext{3}{Department of Physics, Faculty of Science and Engineering, Konan University, 8-9-1 Okamoto, Kobe, Hyogo 658-8501, Japan}
\altaffiltext{4}{Kavli Institute for the Physics and Mathematics of the Universe (WPI), The University of Tokyo Institutes for Advanced Study, The University of Tokyo, 5-1-5 Kashiwa, Chiba 277-8583, Japan}
\altaffiltext{5}{Subaru Telescope, National Astronomical Observatory of Japan, 650 North A'ohoku Place, Hilo, HI 96720, USA}
\altaffiltext{6}{South African Astronomical Observatory, PO Box 9, 7935 Observatory, Cape Town, South Africa}
\altaffiltext{7}{Graduate School of Science and Engineering, Kagoshima University, 1-21-35, Korimoto, Kagoshima, 890-0065, Japan}
\altaffiltext{8}{Department of Astronomy, Kyoto University, Kitashirakawa-Oiwake-cho, Sakyo-ku, Kyoto,  606-8502, Japan}
\altaffiltext{9}{Institute of Astronomy, Graduate School of Science, The University of Tokyo, 2-21-1 Osawa, Mitaka, Tokyo 181-0015, Japan}
\altaffiltext{10}{Institute for Space-Earth Environmental Research, Nagoya University, Furo-cho, Chikusa, Nagoya, Aichi 464-8601, Japan}
\altaffiltext{11}{Institute for Natural and Mathematical Sciences, Massey University, Private Bag 102904 North Shore Mail Centre, Auckland 0745, New Zealand}
\altaffiltext{12}{Department of Earth and Space Science, Graduate School of Science, Osaka University, 1-1 Machikaneyama, Toyonake, Osaka 560-0043, Japan}
\altaffiltext{13}{University of Canterbury, Mt John Observatory, PO Box 56, Lake Tekapo 7945, New Zealand}
\altaffiltext{14}{Nishi-Harima Astronomical Observatory, Center for Astronomy, University of Hyogo, 407-2, Nishigaichi, Sayo, Hyogo 679-5313, Japan}
\altaffiltext{15}{Department of Physics, Tokyo Institute of Technology, 2-12-1 Ookayama, Meguro-ku, Tokyo 152-8551, Japan}
\altaffiltext{16}{Department of Physical Science, Hiroshima University, Kagamiyama, Higashi-Hiroshima 739-8526, Japan}
\altaffiltext{17}{Division of Particle and Astrophysical Science, Graduate School of Science, Nagoya University, Furo-cho, Chikusa-ku, Nagoya, 464-8602, Japan}
\altaffiltext{18}{Department of Physics, Graduate School of Science, Osaka City University, Osaka 558-8585, Japan}
\altaffiltext{19}{Okayama Astrophysical Observatory, National Astronomical Observatory of Japan, 3037-5 Honjo, Kamogata, Asakuchi, Okayama 719-0232, Japan}
\altaffiltext{20}{University of Chinese Academy of Sciences, No.2 Beijing West Road, Purple Mountain Observatory, Nanjing, 210008, China}
\altaffiltext{21}{Precursory Research for Embryonic Science and Technology, Japan Science and Technology Agency, 2-21-1 Osawa, Mitaka, Tokyo 181-0015, Japan}
\altaffiltext{22}{Department of Physics, Toho University, Funabashi, Chiba 274-8510, Japan}
\author{the J-GEM collaboration}

\altaffiltext{$\dagger$}{Deceased 18 August 2017}


\maketitle

\begin{abstract}
The first detected gravitational wave from a neutron star merger was  GW170817.
In this study, we present J-GEM follow-up observations of SSS17a, an electromagnetic counterpart of GW170817.
SSS17a shows a 2.5-mag decline in the $z$-band from 1.7 days to 7.7 days after the merger.
Such a rapid decline is not comparable with supernovae 
% begin ref. com. (MY)
light curves
% end ref. com.
at any epoch.
The color of SSS17a also evolves rapidly and becomes redder
% begin ref. com. (MY) 
for later epochs;
% end ref. com.
the $z-H$ color changed by approximately 2.5 mag in the period of 0.7 days to 7.7 days.
The rapid evolution of both the optical brightness and the color are consistent with the expected properties of a kilonova that is powered by the radioactive decay of newly synthesized $r$-process nuclei.
Kilonova models with Lanthanide elements can reproduce the aforementioned observed properties well, which suggests that $r$-process nucleosynthesis beyond the second peak takes place in SSS17a.
However, the absolute magnitude of SSS17a is brighter than the expected brightness of the kilonova models 
% begin ref. com. (MY) 
% delete ``of''
with the ejecta mass of
% end ref. com.
 0.01 $\Msun$, which suggests a more intense mass ejection ($\sim 0.03 \Msun$) or possibly an additional energy source.
\end{abstract}

\newpage

\section{Introduction}
After the first detections of gravitational wave (GW) events from binary black hole (BBH) coalescence \citep{abbott16,abbott16b,abbott17}, the detection of GWs from a compact binary coalescence including at least one neutron star (NS) has been eagerly awaited. 
This is because the compact binary coalescence including an NS is expected to be accompanied by a variety of electromagnetic (EM) emissions.
An optical and near-infrared (NIR) emission driven by the radioactive decays of $r$-process nuclei, ``kilonova" or ``macronova" \citep{li98,kulkarni05,metzger10}, is one of the most promising EM counterparts.
Optical and NIR observations of these events enable us to understand the origin of $r$-process elements in the Universe as emission properties reflect the ejected mass and  abundances of $r$-process elements \citep[e.g.][]{kasen13,barnes13,tanaka13,tanaka14,metzger14,kasen15}.

\begin{figure}
 \begin{center}
  \includegraphics[width=\linewidth, bb=0 0 299 144]{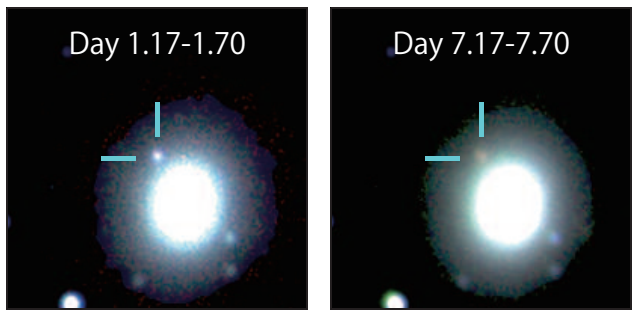}
\end{center}
\caption{Three-color composite images of SSS17a using $z$-, $H$-, and $K_{\rm s}$-band images. The size of the image is 56$\times$56 arcsec$^2$. From left to right, 
% begin ref. com. (MY)
% delete ``the images at $t \sim 1.17-1.70$ and $7.17-7.70$ days after GW170817 was detected are shown.''
the combined images created from the images taken between $t = 1.17$ and $1.70$ days
and between $t = 7.17$ and $7.70$ days are shown.
}\label{fig:image}
\end{figure}

On August 17, 2017, 12:41:04 GMT, the LIGO (Laser Interferometer Gravitational-Wave Observatory) Hanford observatory (LHO) identified a GW candidate in an NS merger \citep{GCN21509}.
The 
% begin ref. com. (MY) 
% delete ``following''
subsequent
% end ref.com.
analysis with three available GW interferometers including the LIGO Livingston Observatory (LLO) and Virgo shrank the localization to 33.6 deg$^2$ for a 90\% credible region \citep{GCN21527} and confirmed the detection \citep[GW170817;][]{GW170817}.
A Fermi-GBM trigger, approximately 2 s after the coalescence, coincided with this GW event and provided additional initial information regarding the localization with an error radius of 17.45 deg \citep{GCN21505}, which 
% begin ref. com. (MY)
% delete ``included the localization from the GWs.''
covers the area localized by the GW detectors.
% end ref. com.
\citet{GCN21529,coulter} reported a possible optical counterpart SSS17a, within the localization area, near NGC 4993.
The source located at $(\alpha, \delta)$=(13:09:48.07, -23:22:53.3), 10 arcsec away from NGC 4993 (Figure \ref{fig:image}), is
an S0 galaxy at a distance of $\sim$40 Mpc \citep{freedman01}. 

We conducted coordinated observations in the framework of Japanese collaboration for Gravitational-wave Electro-Magnetic follow-up (J-GEM) \citep{morokuma16,yoshida17,utsumi17} immediately after the discovery of the strong candidate SSS17a and investigated the characteristics of the optical and NIR emission.
In this paper, we present the results of the J-GEM follow-up observations of SSS17a.
All magnitudes are given using the unit of AB.

\section{J-GEM Observations}
A broad geometrical distribution of observatories was required to observe SSS17a because it was visible for a limited amount of time after sunset in the northern hemisphere.
J-GEM facilities were suitable for observing this target because they are distributed all over the Earth in terms of the longitude, which included the southern hemisphere where the visibility was better.
We used the following facilities to perform follow-up optical observations of GW170817:
8.2 m Subaru / HSC \citep{2012SPIE.8446E..0ZM} and MOIRCS \citep{2008PASJ...60.1347S} at Mauna Kea in the United States;
2.0 m Nayuta / NIC (near-infrared imager) at the Nishi-Harima Astronomical Observatory in Japan;
1.8 m MOA-II 
% begin MY
/ MOA-cam3 
%end MY
\citep{2008ExA....22...51S,2016ApJ...825..112S} and 
%begin MY
% delete ``60''
the 61
%end MY
cm Boller \& Chivens 
% begin MY
telescope
% end MY
(B\&C) 
% begin MY
/ Tripol5
%end MY
at the Mt. John Observatory in New Zealand;
1.5 m Kanata / HONIR \citep{2014SPIE.9147E..4OA} at the Higashi-Hiroshima Astronomical Observatory in Japan;
1.4 m IRSF / SIRIUS \citep{2003SPIE.4841..459N} at the South African Astronomical Observatory; and
50 cm MITSuME \citep{2005NCimC..28..755K} at the Akeno Observatory in Japan.

We reduced all the raw images obtained using the aforementioned instruments in a standard manner.
After eliminating the instrumental signatures, we made astrometric and photometric calibrations.
The astrometric calibrations were performed 
% begin MY
% delete ``using''
with
% end MY
 \emph{astrometry.net} \citep{2010AJ....139.1782L} 
% begin ref. com. (MY)
% delete ``against''
using
%end ref. com.
the default reference catalog USNO-B1.0 \citep{2003AJ....125..984M}, 
while the PanSTARRS catalog \citep{2016arXiv161205560C} was used for the HSC calibration because it is a standard catalog for the HSC reduction.
The number density of stars in the B\&C / Tripol5 images was not sufficient for solving the astrometric solution using \emph{astrometry.net}. We therefore used \emph{Scamp} \citep{2006ASPC..351..112B} for the B\&C / Tripol5 image astrometric calibration instead.
The photometric calibrations were performed using the PanSTARRS catalog for the optical data and 
the 2MASS catalog \citep{2003yCat.2246....0C} for the NIR data.
We did not apply system transformation for adjusting small differences between the band systems because it required the assumption of a spectrum of the source, except in the case of the 
%MOA-II 
MOA-cam3 photometry.  
% begin ref. com. and MY
% delete ``The $R$ (red) band of the MOA-II telescope''
The $R$-band used by MOA-cam3 of MOA-II
%end MY
 is largely different from the standard Johnson system. We determined an empirical relation for the differences between the catalog magnitudes and the instrumental magnitudes as a function of the color 
% begin ref. com. (MY)
% delete ``of''
constructed from
% end ref. com.
 the instrumental magnitudes \citep[e.g.][]{2017AJ....154....3K}. Using Lupton's equation, the catalog magnitudes were converted from the PanSTARRS magnitude to the Johnson system\footnote{http://www.sdss3.org/dr8/algorithms/sdssUBVRITransform.php\#Lupton2005}.
We converted Vega magnitudes to AB magnitudes using the method specified in \citet{2007AJ....133..734B}.

A large contamination from NGC 4993 
% begin MY
% delete ``prevented us from the''
was a problem in performing
% end MY
 accurate measurement of the flux of SSS17a  (Figure \ref{fig:image}).
In order to minimize the systematic uncertainties in the techniques of background subtraction and photometry used for obtaining our measurements, we applied the same procedure for all the data, which is described as follows.
First, we subtracted the host galaxy component from the reduced image using \emph{GALFIT} \citep{galfit}.
The model employed was a Sersic profile with free parameters 
% begin ref. com. (MY)
% delete ``of''
describing 
% end ref. com.
the position, integrated magnitude, effective radius, Sersic index, axis ratio, and position angle. 
A PSF model constructed using \emph{PSFEx} \citep{PSFEx} was used in the fitting procedure.
Once the fitting converged, \emph{GALFIT} generated residual images.
We obtained photometry from the images in which the target was clearly visible after the subtraction.
We then ran \emph{SExtractor 2.19.5} \citep{bertin96} on the residual images, thus
enabling the local sky subtraction with a grid size of 16 pixels---which was larger than the seeing size for all measurements---and the 
% begin MY
PSF
% end MY
model fitting 
%\begin MY
% delete ``capability with a point source.''
for photometry.
% end MY
The residuals of the host galaxy subtraction could be reduced owing to this local sky subtraction.
We adopted  MAG\_POINTSOURCE, an integrated magnitude of the fitted PSF, as a measure of magnitude
and MAGERR\_POINTSOURCE as the error of the measurements.
We confirmed that the measurements
% begin ref. com. and MY
for $z$-band
% end ref. com. and MY
 obtained from \emph{SExtractor} were consistent with those of \emph{hscPipe}, which is a standard pipeline of HSC reduction \citep{2017arXiv170506766B}.
We also confirmed that the brightness of a reference star was constant in all the measurements for an individual instrument.
The measurements are presented in table \ref{tab:data}.
\begin{figure}
 \begin{center}
  \includegraphics[width=\linewidth]{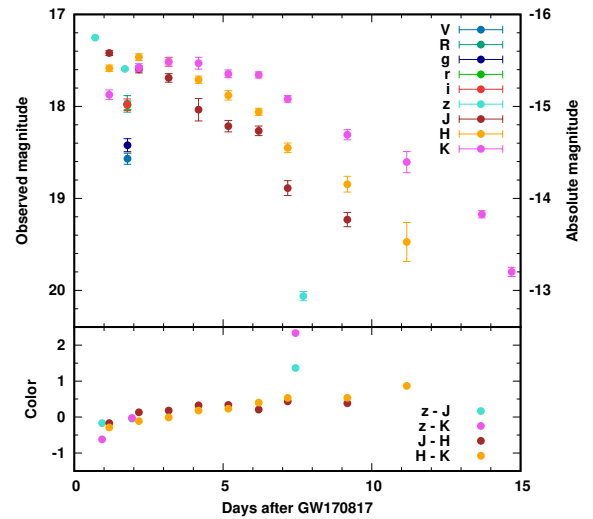}
\end{center}
\caption{Light curves and color evolution of SSS17a. The face color is changed with the respective bands. The galactic extinction has been corrected by assuming $E(B-V) = 0.1$ mag \citep{2011ApJ...737..103S}.}\label{fig:observations}
\end{figure}

\begin{table}[h]
\caption{J-GEM measurements of SSS17a.}\label{tab:data}
\begin{tabular}{ccccc}
\hline
Epoch  & Filter & Mag$^{\ddagger}$  & MagErr & Instrument \\
Days & & [AB] & [AB] &  \\
\hline
0.70 & $z$ & 17.40 & 0.01 & Subaru / HSC \\
1.17 & $J$ & 17.51 & 0.03 & IRSF / SIRIUS \\
1.17 & $H$ & 17.64 & 0.04 & IRSF / SIRIUS \\
1.17 & $K_{\rm s}$ & 17.91 & 0.05 & IRSF / SIRIUS \\
1.70 & $z$ & 17.74 & 0.01 & Subaru / HSC \\
1.78 & $g$ & 18.80 & 0.07 & B\&C / Tripol5 \\
1.78 & $r$ & 18.26 & 0.04 & B\&C / Tripol5 \\
1.78 & $i$ & 18.19 & 0.06 & B\&C / Tripol5 \\
1.78 & $R$ & 18.32$^{*}$ & 0.07 & MOA-II / MOA-cam3 \\
1.79 & $V$ & 18.89$^{*}$ & 0.07 & MOA-II / MOA-cam3 \\
2.17 & $J$ & 17.69 & 0.04 & IRSF / SIRIUS \\
2.17 & $H$ & 17.52 & 0.04 & IRSF / SIRIUS \\
2.17 & $K_{\rm s}$ & 17.61 & 0.04 & IRSF / SIRIUS \\
3.17 & $J$ & 17.78 & 0.05 & IRSF / SIRIUS \\
3.17 & $H$ & 17.57 & 0.04 & IRSF / SIRIUS \\
3.17 & $K_{\rm s}$ & 17.55 & 0.05 & IRSF / SIRIUS \\
4.18 & $J$ & 18.13 & 0.12 & IRSF / SIRIUS \\
4.18 & $H$ & 17.77 & 0.04 & IRSF / SIRIUS \\
4.18 & $K_{\rm s}$ & 17.57 & 0.07 & IRSF / SIRIUS \\
5.18 & $J$ & 18.31 & 0.06 & IRSF / SIRIUS \\
5.18 & $H$ & 17.94 & 0.05 & IRSF / SIRIUS \\
5.18 & $K_{\rm s}$ & 17.68 & 0.04 & IRSF / SIRIUS \\
6.20 & $J$ & 18.36 & 0.05 & IRSF / SIRIUS \\
6.20 & $H$ & 18.12 & 0.04 & IRSF / SIRIUS \\
6.20 & $K_{\rm s}$ & 17.69 & 0.03 & IRSF / SIRIUS \\
7.17 & $J$ & 18.98 & 0.08 & IRSF / SIRIUS \\
7.17 & $H$ & 18.51 & 0.05 & IRSF / SIRIUS \\
7.17 & $K_{\rm s}$ & 17.95 & 0.04 & IRSF / SIRIUS \\
7.70 & $z$ & 20.21 & 0.04 & Subaru / HSC \\
9.18 & $J$ & 19.32 & 0.08 & IRSF / SIRIUS \\
9.18 & $H$ & 18.90 & 0.09 & IRSF / SIRIUS \\
9.18 & $K_{\rm s}$ & 18.34 & 0.06 & IRSF / SIRIUS \\
11.17 & $H$ & 19.53 & 0.21 & IRSF / SIRIUS \\
11.17 & $K_{\rm s}$ & 18.64 & 0.12 & IRSF / SIRIUS \\
14.27 & $K_{\rm s}$ & 19.35 & 0.04 & Subaru / MOIRCS \\
15.27 & $K_{\rm s}$ & 19.97 & 0.05 & Subaru / MOIRCS \\
\hline
\end{tabular}
\begin{tabnote}
\footnotemark[$*$] MOA-II / MOA-cam3 measurements are calibrated using the empirical relation. Uncertainties regarding the conversion are large but not taken into account in the error.
\footnotemark[$\ddagger$] The magnitudes listed here are the values before the Galactic extinction correction.
\end{tabnote}
\end{table}

\section{Results}
% Results
The top panel of Figure \ref{fig:observations} shows the light curves of SSS17a in various bands based on our photometry.
The magnitudes have been corrected for the Galactic extinction by assuming $E(B-V) = 0.10$ mag \citep{2011ApJ...737..103S}.
We do not consider the measurements obtained using Kanata / HONIR, Nayuta / NIC, and MITSuME because the measurements are not reliable owing to the strong contamination from twilight or bad weather. 

A remarkable feature of SSS17a is the rapid decline in the $z$-band brightness 
by 2.5 mag in 6 days. In contrast, the fluxes in the NIR bands decline more slowly 
than those in the optical band. The 
% begin ref. com. (MY)
% delete ``dense''
densely
% end ref. com.
 sampled observations by IRSF / SIRIUS exhibit a 
slight brightening at the earliest epochs in the $H$- and $K_{\rm s}$-bands and 
demonstrate that the light curves in the redder bands start to decline subsequently 
and fade more slowly. The declines in 6 days after the peak are 1.47 mag, 1.33 mag, and 0.96 mag in the $J$-band,
$H$-band, and $K_{\rm s}$-band respectively.

These features are also depicted in the evolution of colors in the $z$-, $J$-, $H$-, 
and $K_{\rm s}$-bands (as shown in the bottom panel of Figure \ref{fig:observations}). The colors 
in the $z$-band and NIR band rapidly become redder, and the reddening in 6 days
% begin ref. com. (MY)
% delete ``is''
are
% end ref. com. 
2.43 mag and 1.00 mag in the $z-K_{\rm s}$ color and $z-J$ color respectively. In contrast, 
the reddening in the colors in the NIR bands 
% begin ref. com. (MY)
% delete ``is''
are
% end ref. com. 
 as slow as 0.34 mag in the $J-H$ color 
and 0.83 mag in the $H-K_{\rm s}$ color.
As a result, the optical-NIR color of SSS17a progressively becomes redder with time (Figure \ref{fig:image}).

\section{Origin of SSS17a}
\begin{figure}[h]
 \begin{center}
    \includegraphics[width=\linewidth]{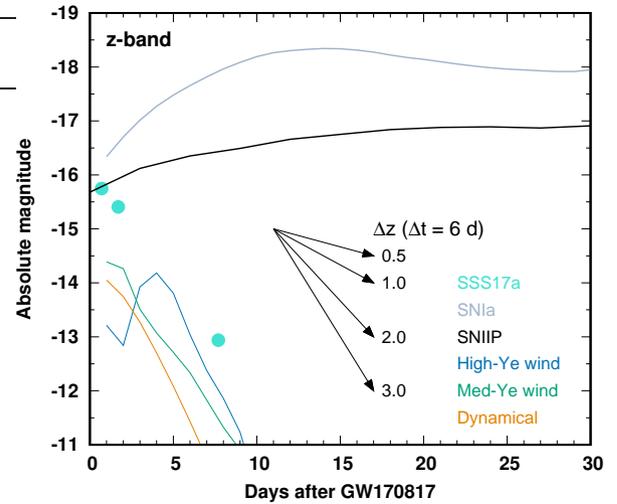} 
 \end{center}
\caption{
Absolute magnitude of $z$-band observations (dots) compared with models of 
% begin MY
% delete ``SNe (in gray)''
supernovae (in gray curves)
% end MY
 and kilonovae (colored curves).
% begin MY
The kilonova models are calculated assuming that the mass of the ejecta from a
neutron star merger $\Mej$ is $0.01 \Msun$.
% end MY
The absolute magnitudes of the kilonova models quickly decline as compared with 
% SNe.
supernovae.
The 
% begin MY
%delete ``observations follow''
$z$-band light curve of SSS17a follows 
%end MY
the decline of the kilonova models although 
% begin MY
% delete ``their brightness is greater by 1--3 magnitudes.''
the observed magnitudes are 1--3 magnitude brighter than the model predictions.
% end MY
The arrows indicate 
% begin MY
%delete ``the decline assuming various $\Delta z$,''
the behaviors of the brightness decline corresponding to various $\Delta z$,
% end MY
 which is the difference in the magnitude of the two epochs for an interval of $\Delta t=6$ days.
}\label{fig:mag_abs}
\end{figure}

\begin{figure}[h]
 \begin{center}
    \includegraphics[width=\linewidth]{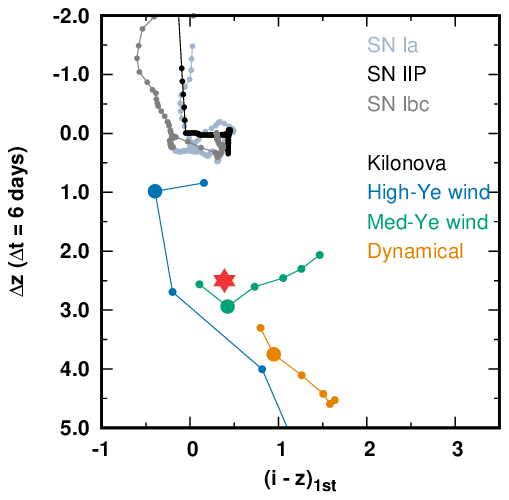} 
 \end{center}
\caption{
% begin MY
% delete "Observation of the $\Delta z$ and $(i-z)_{\rm 1st} plane.''
The result of the photometry of SSS17a is plotted on $\Delta z$ and $(i-z)_{\rm 1st}$ plane
with kilonova and supernova models.
% end MY
For SSS17a (red symbol), $\Delta z$ is the magnitude difference between the two epochs
% begin MY
,
% end MY
$t=1.7$ and 7.7 days ($\Delta t=6$ days) after 
% begin MY
% delete ``the GW170817 is detected,''
the detection of GW170817,
% end MY
 and $(i-z)_{\rm 1st}$ is
the color at the first epoch ($t=1.7$ day).
The models for kilonovae 
% begin MY
% delete ``(colored dots) and supernovae (gray dots) are also shown.''
and supernovae are shown by colored dots and gray dots, respectively.
% end MY
% begin MY
% delete ``A different dot corresponds to a different starting epoch with an increment of 1 day.''
Each dot corresponds to different starting epoch of $\Delta t$ with an increment of 1 day.
% end MY
 The larger dots 
% begin MY
in the kilonova model loci
% end MY
show the values for 
% begin MY
% delete ``the 2nd day from the merger.''
the case that the starting epoch of $\Delta t$ is the 2nd day from the merger.
%end MY 
The kilonova models are located far from the crowds of 
% begin MY
those for 
%end MY
supernovae at 40 Mpc,
especially in terms of $\Delta z$. 
% begin MY
% delete ``The red symbol shows the position of SSS17a, which''
The data point of SSS17a
% end MY
 is consistent with the model of medium $\Ye$ wind.
}\label{fig:dmdc}
\end{figure}
% SSS17a is not similar to supernova

Figure \ref{fig:mag_abs} shows the $z$-band light curves for SSS17a, Type Ia supernova 
\citep[SN Ia,][]{nugent02}, Type II plateau supernova \citep[SN IIP,][]{sanders15}, and 
three kilonova models with an ejecta mass of $\Mej = 0.01 \Msun$ as mentioned by \citet{tanaka17a}. 
The kilonova models 
are a Lanthanide-rich dynamical ejecta model and post-merger wind models with a medium 
$\Ye$ of 0.25 and high $\Ye$ of 0.30. The model with $\Ye =0.25$ contains a small 
fraction of Lanthanide elements while that with $\Ye = 0.30$ is Lanthanide-free.
The rapid decline of SSS17a is not similar to the properties of known supernovae, 
% (SNe),
and the $z$-band magnitude of SSS17a at $t=7.7$ days is $>3$ mag fainter than 
%SNe 
supernovae Ia and IIP.
However, the rapid decline of SSS17a is consistent with the expected 
properties of kilonovae, although SSS17a is 1--3 mag brighter than all the three 
kilonova models. 

The rapid evolution of SSS17a is characterized by a magnitude difference in the 
$z$-band ($\Delta z$) between $t=1.7$ days and 7.7 days (6 days interval).
The red point in Figure \ref{fig:dmdc} shows the $\Delta z$ and $i-z$ color at $t = 1.7$ days \citep[see][]{utsumi17}.
For the purpose of comparison, we show $\Delta z$ in a 6-day interval and $(i-z)_{\rm 1st}$ color at the 1st epoch for 
%SNe 
supernovae
using the spectral template of \citet{nugent02}.
The points show $\Delta z$ and $(i-z)_{\rm 1st}$  with a 1-day step from the day of the merger, and their time evolutions are connected by lines.
The fastest decline 
% begin ref. com. (MY)
% delete ``of''
observed for
% end ref. com.
% SNe 
supernovae is approximately 0.5 mag in 6 days,
and therefore, 
% begin ref. com and MY
% delete ``the model of SNe''
models of supernovae
% end ref. com. and MY
cannot explain the rapid decline of SSS17a.
The wind model with medium $\Ye$ ($\Ye = 0.25$) at $t = 1$ day or 2 days provides the best agreement with the observation.

\begin{figure}
 \begin{center}
    \includegraphics[width=\linewidth]{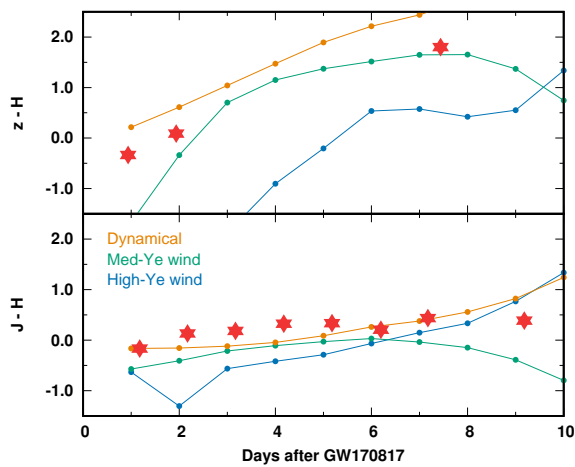} 
 \end{center}
\caption{Time evolution of $z-H$ and $J-H$ color compared with the kilonova models.}\label{fig:color}
\end{figure}
The color evolution of SSS17a is also consistent with that of kilonova models.
Figure \ref{fig:color} shows the $z-H$ and $J-H$ color curves of SSS17a as compared with those of the three kilonova models.
The $J-H$ colors and the absence of the strong evolution are broadly consistent with the models.
The $z-H$ color and its temporal reddening are similar to those of the models comprising Lanthanide elements.
In contrast, the $z-H$ color of SSS17a is not consistent with the Lanthanide-free model (blue curve in Figure \ref{fig:color}), i.e., the high opacities of Lanthanide elements provide a better description of SSS17a.

The properties of SSS17a, i.e., the rapid evolution, red color, and rapid reddening,
are consistent with the standard model of a kilonova.
The color evolution suggests that the ejecta contain a small amount of Lanthanide elements.
This means that $r$-process nucleosynthesis beyond the 2nd peak takes place in the NS merger event GW170817/SSS17a. 
However, the absolute magnitude of the brightness of SSS17a is greater than that of the kilonova models of $\Mej = 0.01 \Msun$.
This discrepancy can be explained by adopting a larger ejecta mass, e.g., $\Mej = 0.03 \Msun$,
which gives a higher radioactive luminosity.
Since the high ejecta mass makes the timescale of the longer, 
a higher ejecta velocity may also be required 
to keep the good agreement in the timescale shown in our paper \citep{tanaka17b}.
Or possibly, a higher luminosity can be accounted for by 
an additional energy source, such as the cocoon emission \citep{2017arXiv170510797G}.

\section{Summary}
We present J-GEM observations of SSS17a, a promising EM counterpart to GW170817.
Intensive observations are performed with Subaru ($z$ and $K_{\rm s}$-band), IRSF ($J$, $H$, and $K_{\rm s}$-band), B\&C ($g$, $r$, and $i$-band), MOA-II ($V$ and $R$-band), Nayuta ($J$, $H$, and $K_{\rm s}$-band), Kanata ($H$-band), and MITSuME ($g$, $r$, and $i$-band) telescopes.
SSS17a exhibits an extremely rapid decline in the $z$-band, which is not explained by any type of supernova at any epoch.
In addition, the evolution of the color is quite rapid; the $z-H$ color is changed by approximately 2.5 mag in 7 days.
We show that the observational properties, i.e., rapid evolution of the light curves, the red color, and its rapid evolution, are consistent with models of kilonovae having Lanthanide elements.
This indicates that $r$-process nucleosynthesis beyond the second peak takes place in the NS merger event GW170817.
However, the absolute magnitude of SSS17a is brighter than that of kilonova models of $\Mej = 0.01 \Msun$. This suggests that the mass ejection is more vigorous ($\sim 0.03 \Msun$) or that there is an additional energy source.

\begin{ack}
We are grateful to the staff of Subaru Telescope, South African Astronomical Observatory,
and University of Canterbury Mt. John Observatory for their help for the observations of
this work.
We thank Dr. Masato Onodera and Prof. Takashi Nakamura who provided insightful comments, and Dr. Nobuhiro Okabe who provided a computational resource.
This work was supported by
MEXT KAKENHI (JP17H06363, JP15H00788, JP24103003, JP10147214, JP10147207) and
JSPS KAKENHI (JP16H02183, JP15H02075, JP15H02069, JP26800103, JP25800103),
the research grant program of the Toyota Foundation (D11-R-0830),
the natural science grant of the Mitsubishi Foundation, 
the research grant of the Yamada Science Foundation,
the National Institute of Natural Sciences (NINS) program for cross-disciplinary science study,
Inoue Science Research Award from Inoue Foundation for Science,
Optical \& Near-Infrared Astronomy Inter-University Cooperation Program from the MEXT, and
the National Research Foundation of South Africa.
This work is based on data collected at the Subaru Telescope, which is operated by the National Astronomical Observatory of Japan, NINS.
%Inter-University Research Program of the Institute for Cosmic Ray Research, the University of Tokyo.

\end{ack}

\bibliographystyle{myaasjournal}

\bibliography{reference}

\end{document}